\documentclass[%
 aip,%
 apl
 cp,%
 amsmath,amssymb,%
 reprint,%
]{revtex4-1}

\usepackage{graphicx}% Include figure files
\usepackage{dcolumn}% Align table columns on decimal point
\usepackage{bm}% bold math
\usepackage{makecell}
\begin{document}

\preprint{AIP/123-QED}

\title{Concentric transmon qubit featuring fast tunability and an anisotropic magnetic dipole moment}% Force line breaks with \\
%\thanks{Footnote to title of article.}

\author{Jochen Braum\"uller}
\email{jochen.braumueller@kit.edu.}
\affiliation{Physikalisches Institut, Karlsruhe Institute of Technology, 76131 Karlsruhe, Germany}

\author{Martin Sandberg}
\author{Michael R. Vissers}
\affiliation{National Institute of Standards and Technology, Boulder, Colorado 80305, USA}

\author{Andre Schneider}
\author{Steffen Schl\"or}
\author{Lukas Gr\"unhaupt}
\author{Hannes Rotzinger}
\affiliation{Physikalisches Institut, Karlsruhe Institute of Technology, 76131 Karlsruhe, Germany}

\author{Michael Marthaler}
\affiliation{Institut f\"ur Theoretische Festk\"orperphysik, Karlsruhe Institute of Technology, 76131 Karlsruhe, Germany}

\author{Alexander Lukashenko}
\author{Amadeus Dieter}
\affiliation{Physikalisches Institut, Karlsruhe Institute of Technology, 76131 Karlsruhe, Germany}

\author{Alexey V. Ustinov}
\affiliation{Physikalisches Institut, Karlsruhe Institute of Technology, 76131 Karlsruhe, Germany}
\affiliation{National University of Science and Technology MISIS, Moscow 119049, Russia}

\author{Martin Weides}
\affiliation{Physikalisches Institut, Karlsruhe Institute of Technology, 76131 Karlsruhe, Germany}
\affiliation{Physikalisches Institut, Johannes Gutenberg University Mainz, 55128 Mainz, Germany}

\author{David P. Pappas}
\affiliation{National Institute of Standards and Technology, Boulder, Colorado 80305, USA}

\date{\today}% It is always \today, today,

\begin{abstract}
We present a planar qubit design based on a superconducting circuit that we call concentric transmon. While employing a straightforward fabrication process using Al evaporation and lift-off lithography, we observe qubit lifetimes and coherence times in the order of $10\,\mathrm{\mu s}$. We systematically characterize loss channels such as incoherent dielectric loss, Purcell decay and radiative losses. The implementation of a gradiometric SQUID loop allows for a fast tuning of the qubit transition frequency and therefore for full tomographic control of the quantum circuit. Due to the large loop size, the presented qubit architecture features a strongly increased magnetic dipole moment as compared to conventional transmon designs. This renders the concentric transmon a promising candidate to establish a site-selective passive direct $\hat Z$ coupling between neighboring qubits, being a pending quest in the field of quantum simulation.
\end{abstract}

%\pacs{Valid PACS appear here}

\maketitle

%\section{\label{sec:introduction}Introduction}

%\textit{Introduction}
Quantum bits based on superconducting circuits are leading candidates for constituting the basic building block of a prospective quantum computer. A common element of all superconducting qubits is the Josephson junction. The nonlinearity of Josephson junctions generates an anharmonic energy spectrum in which the two lowest energy states can be used as the computational basis \cite{Clarke2008,Schoelkopf2008}. Over the last decade there has been a two order of magnitude increase in coherence times of superconducting qubits. This tremendous improvement allowed for demonstration of several major milestones in the pursuit of scalable quantum computing, such as the control and entanglement of multiple qubits \cite{Steffen2006,Barends2014}. Further increases in coherence times will eventually allow for building a fault tolerant quantum computer with a reasonable overhead in terms of error correction, as well as implementing novel quantum simulation schemes by accessing wider experimental parameter ranges \cite{Paraoanu2014}. While superconducting qubits embedded in a 3D cavity \cite{Paik2011} have shown coherence times in excess of $100\,\mathrm{\mu s}$ \cite{Rigetti2012}, this approach may impose some constraint on the scalability of quantum circuits. Since the Josephson junction itself does not limit qubit coherence \cite{Paik2011}, comparably long lifetimes can also be achieved in a planar geometry by careful circuit engineering.

\begin{figure} %++++++++++++++++++++++++++++++++++++++FIG1++++++++++++++++++++++++++++++++++++++++++
%\includesvg{sample07}
\includegraphics{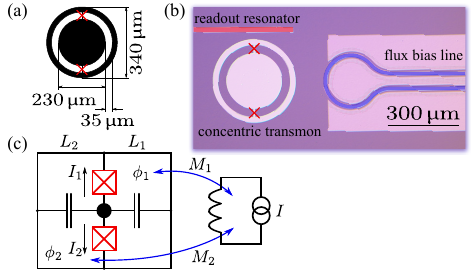}
\caption{(a) Geometry of the concentric transmon qubit. Two Josephson junctions (red crosses), located opposite to each other, connect the central island to an outer ring. (b) Optical micrograph of the fabricated sample. The readout resonator (red) capacitively couples to the concentric transmon from above. The on-chip flux bias line (blue) is visible to the right. It is designed in coplanar geometry, having the microwave ground reference in the device plane (bright color). The flux bias line is grounded at one end on the chip. (c) Schematic circuit diagram of the concentric transmon, revealing the gradiometric SQUID architecture. The central part, marked with a black dot, corresponds to the center island of the transmon. Since the mutual inductances to the flux bias line are not equal, $M_1\neq M_2$, the effective critical current of the SQUID can be tuned.}
\label{fig:ring_transmon}
\end{figure}

In this paper, we present the design and characterization of a superconducting quantum circuit comprising a concentric transmon qubit \cite{Sandberg2015}, schematically depicted in Fig.~\ref{fig:ring_transmon}(a). The two capacitor pads forming the transmon's large shunt capacitance are implemented by a central disk island and a concentrically surrounding ring. The two islands are interconnected by two Josephson junctions forming a gradiometric SQUID. A $50\,\mathrm{\Omega}$ impedance matched on-chip flux bias line located next to the qubit allows for fast flux tuning of the qubit frequency due to the imposed asymmetry. This guarantees high experimental flexibility and enables full tomographic control. The gradiometric flux loop design reduces the sensitivity to external uniform magnetic fields and thus to external flux noise. For readout and control purposes we embed the qubit in a microstrip resonator circuit, forming a familiar circuit quantum electrodynamics (cQED) system \cite{Wallraff2004}. The sample is fabricated in hybrid coplanar-microstrip geometry, featuring a ground plane on the backside of the substrate, thus exploiting the increase in mode volume in a microstrip architecture. The circular shape of the transmon features a strongly reduced electric dipole moment due to symmetry and hence a reduction in radiation loss can be expected \cite{Sandberg2013} compared to a regular pad geometry of equal size and electrode spacing. The sample is prepared in a straightforward fabrication process using pure aluminum for the metalization and employing the conventional shadow angle evaporation technique. The best measured relaxation and decoherence times are in the order of $10\,\mathrm{\mu s}.$

%\section{\label{sec:theory}Theory/Design}

%flux bias loop
%\textit{Design}
The schematic circuit diagram of the concentric transmon is depicted in Fig.~\ref{fig:ring_transmon}(c). The two Josephson junctions are connecting the central island to the outer ring of the transmon. The islands act as coplanar electrodes \cite{Sandberg2013} giving rise to the total qubit capacitance $C=81\,\mathrm{fF}$, including the contribution by the ground plane and coupling capacitances. Considering loop $L_1$ as the primary transmon loop, the gradiometric dc-SQUID architecture can be recognized. The kinetic inductance of the superconducting aluminum wire can be neglected due to its thickness and the width of the Josephson junction leads. The effective critical current $I_{c,eff}$ of the SQUID is tuned by applying an inhomogeneous magnetic field supplied by the on-chip flux bias line. Due to the gradiometric geometry, the effective critical current $I_{c,eff}=2I_c|\cos(\pi\Delta\Phi/2\Phi_0)|$ in the primary loop $L_1$ is $2\Phi_0$-periodic in the flux asymmetry $\Delta\Phi=|\Phi_1-\Phi_2|$ between the loops. Here, $\Phi_0=h/2e$ denotes the magnetic flux quantum. We analytically calculate the net mutual inductance to the flux bias line to be $2.3\,\mathrm{pH}$. This yields a flux bias current of $0.9\,\mathrm{mA}$ required to induce a $\Phi_0$ in the primary transmon loop.

%\textit{Fabrication}
The fabrication process consists of two subsequent deposition and lithography steps. The feedline, the microstrip resonator and the flux bias line are structured by optical lithography in a lift-off process of a $50\,\mathrm{nm}$ thick aluminum film. After that, the concentric transmon including the Josephson junctions are patterned using electron beam lithography. The Josephson junctions are formed by shadow angle evaporation with electrode film thicknesses of $30\,\mathrm{nm}$ and $50\,\mathrm{nm}$, respectively. A $50\,\mathrm{nm}$ thick aluminum film is applied on the backside of the intrinsic silicon substrate. Both metalizations on the device side of the chip are deposited by aluminum evaporation at a background chamber pressure of about $3\cdot 10^{-8}\,\mathrm{mbar}$. A micrograph of the fabricated sample is depicted in Fig.~\ref{fig:ring_transmon}(b).

%parameters
%\textit{Parameters}
We operate the device at a fundamental qubit transition frequency $\omega_q/2\pi =6.85\,\mathrm{GHz}$, away from any flux sweet spot and far detuned from the readout resonator to reduce Purcell dissipation. The Josephson energy $E_J/h=29\,\mathrm{GHz}$ dominates the charging energy $E_C/h=0.24\,\mathrm{GHz}$, yielding $E_J/E_C=120$. This assures the circuit to be inherently insensitive to charge noise and features an intrinsic self-biasing \cite{Koch2007}. The total critical current of both Josephson junctions is $58\,\mathrm{nA}$. Due to its large loop size, the concentric transmon exhibits a notable geometric inductance $L_g=0.6\,\mathrm{nH}$, accounting for $10\%$ of the total qubit inductance. As one important consequence, the qubit anharmonicity is not constant with respect to the bias point. For the operation point we extract a relative anharmonicity of $3.4\%$. Relevant qubit parameters are extracted by fitting an effective system Hamiltonian to spectroscopic data, see Supplemental Material \cite{Supp2015}. By sweeping the flux bias current we observe a period of $1.7\,\mathrm{mA}$ in the qubit transition frequency, confirming the expected $2\Phi_0$-periodicity in good approximation.

\begingroup
\squeezetable
\begin{table} %++++++++++++++++++++++++++++++++++++++TABLE1++++++++++++++++++++++++++++++++++++++++++
\begin{center}
\caption{Calculated loss contributions fo the concentric transmon qubit. The main contribution $\Gamma_{1,ind} ^{-1}$ arises from inductive coupling to the flux bias line, leading to a total Purcell limitation of $\Gamma_{1,P}^{-1}=16\,\mathrm{\mu s}$. The estimated reciprocal sum $\Gamma_{\Sigma} ^{-1}$ is in good agreement with the measured $T_1$.}
\label{tab:T1limitations}
%\vspace*{1mm}
\begin{ruledtabular}
\begin{tabular}{ccccc|c}
%\begin{tabular}{p{1.35cm}p{1.4cm}p{1.4cm}cp{1.15cm}p{1.15cm}|p{1.3cm}}
\multicolumn{3}{c}{Purcell} & defects & radiation & reciprocal sum\\
%\cline{1-3}

%sweet spot data: Purcell: 47.3us 32.2us 87.2us -> 15.8us; TLF: 25.8us, rad 100us -> total: 3.7us
$\Gamma_{1,sm} ^{-1}$ & $\Gamma_{1,ind} ^{-1}$ & $\Gamma_{1,cap} ^{-1}$ & $\Gamma_{1,TLF} ^{-1}$ & $\Gamma_{1,rad}^{-1}$ & $\Gamma_{\Sigma} ^{-1}$\vspace*{1mm}\\
\hline

\vline height8pt width0pt\relax$47\,\mathrm{\mu s}$ & $32\,\mathrm{\mu s}$ & $\sim 87\,\mathrm{\mu s}$ &&&\\

%\cline{1-3}
\multicolumn{3}{c}{$\Gamma_{1,P}^{-1}=16\,\mathrm{\mu s}$} & $\sim 26\,\mathrm{\mu s}$ & $\gtrsim 100\,\mathrm{\mu s}$ & $8.9\,\mathrm{\mu s}$ \vline height8pt width0pt\relax\\

%sweet spot data: Purcell: 14us 8us 69us -> 4.7us; TLF: 22us, rad 100us -> total: 3.7us

\end{tabular}
\end{ruledtabular}
\end{center}
\end{table}
\endgroup

%relaxation time introduction
%\textit{Loss mechanisms}
Since coherence is relaxation limited, $T_2\leq 2T_1$, taking effort in increasing $T_1$ typically results in an increase in $T_2$. To accomplish this, we engineered the concentric transmon to have a reduced sensitivity to its major loss channels, namely spontaneous Purcell emission, radiative dipole decay and loss due to surface defect states. Losses due to quasiparticle tunneling processes typically impose a $T_1$ limitation at around $\sim 1\,\mathrm{ms}$ \cite{Riste2013}, having no considerable effect on the lifetime of our circuit. Relevant contributions leading to qubit decay are summarized in Tab.~\ref{tab:T1limitations}.

%Purcell
%\textit{Purcell decay}
The coupling limited quality factor $Q_L=\omega_r/\kappa$ of the dispersive readout resonator at $\omega_r/2\pi =8.79\,\mathrm{GHz}$ is $2.1\cdot 10^3$, close to the design value. The qubit lifetime is potentially Purcell limited by spontaneous emission into modes that are nearby in frequency. Major contributions are the dispersive single-mode decay into the capacitively coupled readout resonator as well as emission into the flux bias line \cite{Koch2007,Houck2008} due to inductive coupling. The coupling constant $g/2\pi=55\,\mathrm{MHz}$ between qubit and resonator is extracted from the dispersive shift of the resonator \cite{Braumuller2015}. We calculate a single-mode Purcell limitation induced by the readout resonator of $\Gamma_{1,sm}=\kappa\left(g/\Delta\right)^2=1/47\,\mathrm{\mu s}$ at a detuning $\Delta =1.94\,\mathrm{GHz}$ from the readout resonator. The multi-mode Purcell limitation due to inductive coupling of the qubit to the flux bias line of impedance $Z_0=50\,\mathrm{\Omega}$ is $\Gamma_{1,ind}=\omega_q ^2 M^2/(LZ_0)=1/32\,\mathrm{\mu s}$, with $M=2.3\,\mathrm{pH}$ being the gradiometric mutual inductance and $L=6.3\,\mathrm{nH}$ the total qubit inductance. This rather stringent loss channel can be reduced further by decreasing the coupling to the flux bias line. By an analogous approach to account for Purcell decay due to capacitive coupling to the flux bias line ($C_c\sim 0.1\,\mathrm{fF}$), we calculate $\Gamma_{1,cap}=\omega_q ^2 C_c ^2 Z_0/C\sim 1/87\,\mathrm{\mu s}$. The presented values are approximations and give a rough estimate of the expected Purcell loss. A more stringent analysis would require a full 3D electromagnetic simulation of the circuit and a blackbox quantization \cite{Solgun2014}.

%microstrip, defects
%\textit{Dielectric defect loss}
The coplanar concentric transmon is embedded in a microstrip geometry, where the ground reference on the device side of the substrate is substituted by a backside metalization. The largest fraction of the electric field energy is stored in the substrate, and the field strength at incoherent and weakly coupled defects residing in surface and interface oxides of the sample is reduced due to an increased mode volume. Highest fields in the geometry appear in the gap between the center island and the ring of the concentric transmon within the substrate. From the vacuum energy of the transmon we extract a weighted mean electrical field strength $|\vec E|=2.3\,\mathrm{V/m}$ in the surface and interface oxide of an estimated effective volume of $V=50\cdot 10^{-18}\,\mathrm{m^3}$. We assume a maximum defect dipole moment $|\vec d _0|=1.6\,\mathrm{e\r{A}}$, reported in literature \cite{Martinis2005,Cole2010} as the highest dipole moment observed in Josephson junction barriers and therefore yielding a worst case estimation. We employ a normalized dipole moment distribution \cite{Martinis2005} $P(p)=A\sqrt{1-p^2}/p$, with relative dipole moment $p=|\vec{d}|/|\vec d _0|$. Taking into account a normalized defect probability distribution \cite{Muller2015} $P(\omega,\theta)=B\omega^\alpha \cos^\alpha \theta/\sin\theta$ one can estimate the mean relaxation rate $\Gamma_{1,TLF}$ due to a single incoherent two level fluctuator (TLF) with averaged parameters to be
\begin{equation}
%\langle\Gamma_{1,TLF}\rangle _1=
\int_0 ^{d_0}\mathrm{d}pP(p)\frac{p^2|\vec E|^2}{\hbar^2}\int_0^{\omega_{TLF}}\mathrm{d}\omega\int_0^{\pi/2}\mathrm{d}\theta P(\omega,\theta)C(\omega_q).
\label{eq:Gdef}
\end{equation}
$\omega$ denotes the TLF frequency that we integrate to a maximum frequency $\omega_{TLF}/2\pi=15\,\mathrm{GHz}$, and $\theta$ sets the dipole matrix element $\sin\theta$. The spectral density $C(\omega_q)=\sin^2\theta\,\gamma_2/(\gamma_2^2+(\omega -\omega_q)^2)$ essentially is the Fourier transform of the coupling correlation function \cite{Muller2015}, with a TLF dephasing rate $\gamma_2/2\pi=10\,\mathrm{MHz}$. The averaged rate induced by a single TLF, given in Eq.~\eqref{eq:Gdef}, is multiplied by the number $N=\rho_0V\hbar\omega_{TLF}$ of defect TLF interacting with the qubit. With a distribution parameter $\alpha=0.3$ \cite{Faoro2015} and a constructed defect density $\rho_0=4\cdot 10^2/\mathrm{\mu m^3}/\mathrm{GHz}$ we compute a $T_1$ limitation due to the bath of incoherent TLF to be $\Gamma_{1,TLF}\sim 1/26\,\mathrm{\mu s}$. The choice of $\rho_0$ is consistent with literature \cite{Barends2013,Martinis2005} and is justified by a very good agreement with a loss participation ratio analysis carried out via a finite element simulation, see Supplemental Material \cite{Supp2015}. The calculated decay rate shows a very weak dependence on the employed parameters $\gamma_2$, $\alpha$ and integration cutoffs. $\Gamma_{1,TLF}$ imposes a limitation for the best measured $T_1$ due to the quasi-static bath of incoherent defects. In general, defects in the Josephson junction of the device do not significantly contribute to qubit decay. Due to its small size, the defect density in the Josephson barrier is discretized and therefore highly reduced \cite{Weides2011,Weides_Transmon}.

%radiation
%\textit{Radiation loss}
As pointed out in Ref.~[\onlinecite{Sandberg2013}], radiative loss becomes apparent for qubits with a large electric dipole moment. Radiative decay is reduced as the dipole of the mirror image, induced by the ground plane of the microstrip geometry, radiates in antiphase, leading to destructive interference. Our circular geometry brings about an additional decrease in radiated power by strongly reducing the electric dipole moment of the qubit. We analyze this by simulating the dissipated power in a conductive plane placed $1.5\,\mathrm{mm}$ above our geometry in the medium far field and compare the result to a conventional pad architecture. The internal quality factor of the qubit eigenmode indicates a radiative contribution of $\Gamma_{1,rad}^{-1}\gtrsim 100\,\mathrm{\mu s}$. The radiative dissipation of a comparable qubit in pad geometry exceeds this value by more than an order of magnitude.

%\section{\label{sec:experiment}Experimental methods}

%\section{\label{sec:results}Results and analysis}

%\subsection{Energy relaxation, qubit coherence}

\begin{figure} %++++++++++++++++++++++++++++++++++++++FIG2++++++++++++++++++++++++++++++++++++++++++
%\includesvg{decay08}
\includegraphics{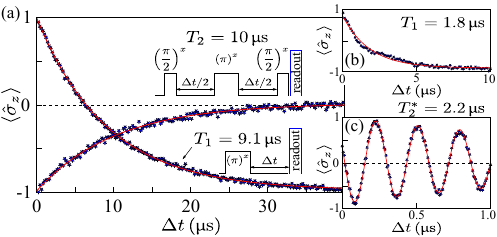}
\caption{(a) Measured relaxation time $T_1=9.1\,\mathrm{\mu s}$ and dephasing time $T_2=10\,\mathrm{\mu s}$ with applied measurement pulse sequences shown in the inset. The operation frequency is $\omega_q/2\pi =6.85\,\mathrm{GHz}$, corresponding to a detuning of $1.9\,\mathrm{GHz}$ below the readout resonator frequency. A Hahn echo pulse is applied in the Ramsey sequence to compensate for low-frequency fluctuations in the qubit transition frequency. (b), (c) $T_1$ decay and bare dephasing without echo pulse $T_2^*$ measured at the flux sweet spot at $7.72\,\mathrm{GHz}$. While $T_1$ is reduced due to Purcell dissipation, $T_2^*$ is comparable to the bare $T_2^*$, measured far detuned from the sweet spot. The slight deviation from an exponential decay is attributed to excitation leakage.}
\label{fig:decay}
\end{figure}

%\textit{Dissipative dynamics}
The dissipative dynamics of the investigated concentric transmon is depicted in Fig.~\ref{fig:decay}. We excite the qubit by applying a $\pi$ pulse and measure its state $\langle\hat\sigma _z\rangle$ with a strong readout pulse after a varying time delay $\Delta t$. The obtained energy relaxation time $T_1=9.1\,\mathrm{\mu s}$ in Fig.~\ref{fig:decay}(a), lower part, is in good agreement with the estimated $T_1$ time, see Tab.~\ref{tab:T1limitations}. To verify the presented loss participation ratio analysis, $T_1$ decay close to the flux sweet spot is shown in Fig.~\ref{fig:decay}(b). The Purcell contribution is calculated to be $4.7\,\mathrm{\mu s}$, reducing the overall dissipation estimate to $3.7\,\mathrm{\mu s}$. The anharmonicity at the flux sweet spot is reduced to $1\%$, favoring leakage into higher levels and therefore degrading the reliability of the exponential fit. The reduction in $T_1$ at the sweet spot is not fully understood. A possible explanation is that the presented first order model for estimating Purcell decay is not very accurate for this high geometric inductance transmon. In addition, losses related to the geometric inductance such as quasi particles or magnetic vortices may be more prominent at the flux sweet spot.

In continuous lifetime measurements, we observe temporal fluctuations in $T_1$ down to about $2\,\mathrm{\mu s}$. We conjecture that this is attributed to discrete TLF dynamics of individual TLF, located in the small oxide volume at the leads to the Josephson junctions, where the electric field strength is enhanced.

The spin echo dephasing time is $T_2=10\,\mathrm{\mu s}$, see Fig.~\ref{fig:decay}(a), upper part. Due to the Hahn echo pulse in between the projecting $\pi/2$ pulses, our device is insensitive to low-frequency noise roughly below $1/\Delta t=25\,\mathrm{kHz}$. A Ramsey $T_2^* \sim 2\,\mathrm{\mu s}$ is measured without echo pulse. It is interesting to note that a rather high $T_2$ is maintained in spite of the large loop size and while operating the transmon detuned by $1\,\mathrm{GHz}$ from its flux sweet spot, where flux noise contributes to pure dephasing. Figure~\ref{fig:decay}(c) shows the dephasing time $T_2^*$ without echo pulse measured at the flux sweet spot. Dephasing, presumably induced in part by local magnetic fluctuators, is reduced due to the vanishing slope of the energy dispersion at the flux sweet spot. In a non-tunable TiN version of the concentric transmon with a single Josephson junction and an opening in the outer ring, we find maximal $T_1\sim 50\,\mathrm{\mu s}$ and dephasing times up to $T_2 ^* \sim 60\,\mathrm{\mu s}$ without Hahn echo. Similar coherence times might be achieved for the tunable concentric design by implementing it in a TiN material system. The measured coherence times compare with other planar transmon geometries such as a non-tunable Al based transmon with decreased finger gap size \cite{Chow2012} ($T_1=9.7\,\mathrm{\mu s}$, $T_2 ^* =10\,\mathrm{\mu s}$) and the cross shaped transmon \cite{Barends2013} ($T_1=40\,\mathrm{\mu s}$, $T_2=20\,\mathrm{\mu s}$).

\begin{figure} %++++++++++++++++++++++++++++++++++++++FIG3++++++++++++++++++++++++++++++++++++++++++
%\includesvg{tune04}
\includegraphics{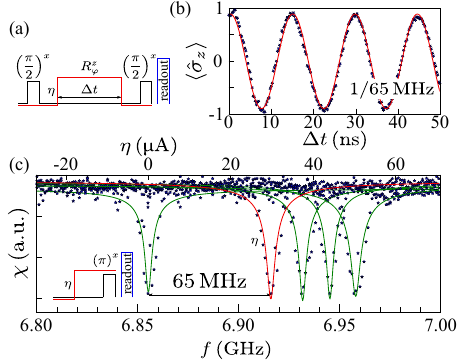}
\caption{Demonstration of fast $\hat Z$ tunability of the concentric transmon. (a) Pulse sequence applied for calibration. Between two projecting $\pi /2$-pulses we apply a $\hat Z$-rotation $R_\varphi ^z$ of amplitude $\eta$ and length $\Delta t$. (b) The expectation value of the qubit state oscillates between its fundamental states $|0\rangle$, $|1\rangle$ with a relative Larmor frequency $\omega_L/2\pi=65\,\mathrm{MHz}$ for a pulse amplitude of $\eta=32\,\mathrm{\mu A}$. (c) Check measurement of the bare and detuned qubit transition frequency $\omega_q$ by exciting with a $\pi$-pulse (see inset). The initial frequency is shifted dependent on the amplitude $\eta$ of the applied current pulse, see different Lorentzian fits. For $\eta=32\,\mathrm{\mu A}$, as applied in (a), we obtain a frequency shift in accordance with the one extracted in (b), see red Lorentzian fit.}
\label{fig:tune}
\end{figure}

Figure~\ref{fig:tune} demonstrates fast frequency control of the concentric transmon. This is commonly referred to as $\hat Z$ control since Pauli's spin operator $\hat\sigma _z$ appears in the Hamiltonian representation of the qubit. We record the equatorial precession (see Fig.~\ref{fig:tune}(b)) due to a detuning pulse $R_\varphi ^z$ of amplitude $\eta$ in between two projective $\pi/2$ pulses. The pulse sequence is given in Fig.~\ref{fig:tune}(a). The pulse amplitude $\eta$ translates into magnetic flux applied to the flux bias line, see the experimental setup in the Supplemental Material \cite{Supp2015}. In the laboratory frame, this corresponds to a shift in qubit frequency $\Delta f\propto \eta$, which is confirmed in a quasi-spectroscopic measurement by exciting the detuned qubit with a $\pi$ pulse, see Fig.~\ref{fig:tune}(c). In the frame rotating with the qubit frequency the $R_\varphi ^z$ pulse induces a change in Larmor frequency, leading to an effective precession in this rotating frame with an angular frequency proportional to the detuning pulse amplitude $\eta$. By increasing $\eta$, we demonstrated a fast frequency detuning of up to $200\,\mathrm{MHz}$ (not shown here). A further increase of the tunability range, requiring pulse shaping, renders our device a valuable tool for a variety of quantum experiments.

%\textit{Site-selective $\hat Z$ coupling}
We propose the concentric transmon as a suitable candidate to establish a direct inductive $\hat Z$ coupling between neighboring qubits. Since the area of the magnetic flux loop is large, the magnetic dipole moment of our qubit is strongly enhanced as compared to conventional transmon designs where it is typically negligible. For two concentric transmon qubits separated by $50\,\mathrm{\mu m}$, we estimate an inductive $\hat Z$ coupling in the range of $g_z ^{ind}/2\pi\sim 1\,\mathrm{MHz}$ for an operation point where the flux dependent qubit spectrum has a large slope \cite{Supp2017}. The mutual inductance is calculated by applying the double integral Neumann formula \cite{Jackson1962}. At the given qubit distance, the capacitive coupling is simulated to be in the order of $15\,\mathrm{MHz}$. To further increase the $\hat Z$ coupling, the qubit geometry can be adapted to enhance the mutual inductance. Especially for neighboring qubits being detuned in frequency, the $\hat Z$ coupling may dominate the effective transversal capacitive coupling. Since the effective mutual inductance vanishes completely for adjacent qubits arranged at a relative rotation angle of $\pi/2$ and is altered when their Josephson junctions are aligned, the concentric transmon geometry allows for a site-selective $\hat Z$ coupling. We consider this scheme as highly promising for the field of quantum simulation, for instance in the context of implementing Ising spin models. Patterning an array of concentric transmon qubits featuring $\hat X$ and $\hat Z$ coupling along two orthogonal directions while exploiting the strongly reduced off-site inductive interaction may be a route to implement a quantum neural network, which is a powerful tool in quantum computation \cite{Schuld2014}.

%\section{\label{sec:conclusion}Conclusion}

%\textit{Conclusion}
In conclusion, we introduced a planar tunable qubit design based on a superconducting circuit that we call concentric transmon. The observed qubit lifetimes and coherence times are in the order of $10\,\mathrm{\mu s}$ and thereby competitive with conventional transmon geometries. The qubit lifetime is Purcell limited by its readout resonator and the flux bias line. Radiative loss, which is an intrinsic loss channel to the geometry, is demonstrated to be limiting only above $\sim 100\,\mathrm{\mu s}$, evincing the potential of the reported architecture. A major advantage of our approach is the straightforward fabrication process. We demonstrated full tomographic control of our quantum circuit and discuss the high potential of the presented qubit design for the implementation of a direct site-selective $\hat Z$ coupling between neighboring qubits, being a pending quest in quantum simulation.

\begin{acknowledgments}
The authors are grateful for valuable discussions about two-level defects with C. M\"uller and J. Lisenfeld. We want to thank L. Radtke for assistance in sample fabrication and S. T. Skacel for providing microwave simulations.
This work was supported by the European Research Council (ERC) within consolidator Grant No. 648011, Deutsche Forschungsgemeinschaft (DFG) within project No. WE4359/7-1, and through the DFG-Center for Functional Nanostructures National Service Laboratory (CFN-NSL). This work was also supported in part by the Ministry for Education and Science of the Russian Federation Grant No. 11.G34.31.0062 and by NUST MISIS under Contract No. K2-2014-025.
J.B. acknowledges financial support by the Helmholtz International Research School for Teratronics (HIRST) and the Landesgraduiertenf\"orderung (LGF) of the federal state Baden-W\"urttemberg.

M.S. conceived the concentric transmon geometry. J.B. designed, fabricated and measured the tunable concentric transmon circuit. M.S. and M.R.V. designed and fabricated the TiN transmon.
\end{acknowledgments}

%\nocite{*}
\bibliography{concentric}

\end{document}

\preprint{AIP/123-QED}

\title{Concentric transmon qubit featuring fast tunability and an anisotropic magnetic dipole moment - Supplemental Material}% Force line breaks with \\
%\thanks{Footnote to title of article.}

\author{Jochen Braum\"uller}
\email{jochen.braumueller@kit.edu.}
\affiliation{Physikalisches Institut, Karlsruhe Institute of Technology, 76131 Karlsruhe, Germany}

\author{Martin Sandberg}
\author{Michael R. Vissers}
\affiliation{National Institute of Standards and Technology, Boulder, Colorado 80305, USA}

\author{Andre Schneider}
\author{Steffen Schl\"or}
\author{Lukas Gr\"unhaupt}
\author{Hannes Rotzinger}
\affiliation{Physikalisches Institut, Karlsruhe Institute of Technology, 76131 Karlsruhe, Germany}

\author{Michael Marthaler}
\affiliation{Institut f\"ur Theoretische Festk\"orperphysik, Karlsruhe Institute of Technology, 76131 Karlsruhe, Germany}

\author{Alexander Lukashenko}
\author{Amadeus Dieter}
\affiliation{Physikalisches Institut, Karlsruhe Institute of Technology, 76131 Karlsruhe, Germany}

\author{Alexey V. Ustinov}
\affiliation{Physikalisches Institut, Karlsruhe Institute of Technology, 76131 Karlsruhe, Germany}
\affiliation{National University of Science and Technology MISIS, Moscow 119049, Russia}

\author{Martin Weides}
\affiliation{Physikalisches Institut, Karlsruhe Institute of Technology, 76131 Karlsruhe, Germany}
\affiliation{Physikalisches Institut, Johannes Gutenberg University Mainz, 55128 Mainz, Germany}

\author{David P. Pappas}
\affiliation{National Institute of Standards and Technology, Boulder, Colorado 80305, USA}

\date{\today}% It is always \today, today,
\maketitle

\section{Experimental Setup}

\begin{figure}[h] %++++++++++++++++++++++++++++++++++++++FIG1++++++++++++++++++++++++++++++++++++++++++
\includegraphics{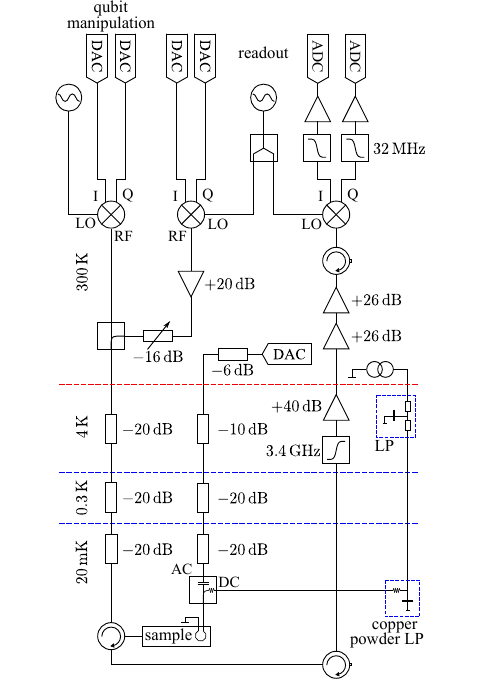}
%\includesvg{setup02}
\caption{Schematic diagram of the experimental setup.}
\label{fig:setup}
\end{figure}

The quantum circuit is mounted in an aluminum box and is cooled down to about $20\,\mathrm{mK}$ in a dilution refrigerator. Its sample box is enclosed in a cryoperm case for additional magnetic shielding. The base stage of the cryostat is surrounded by a radiation shield having its inner surface coated with blackbody-absorbing material to shield the sample from stray infrared radiation generating quasiparticles \cite{Barends2011}.

The schematic diagram of the measurement and microwave setup is depicted in Fig.~\ref{fig:setup}. To perform measurements on the concentric transmon qubit we use a pulsed readout tone. The resonator shift is extracted from the microwave reflection signal at a single-ended $50\,\mathrm{\Omega}$ matched transmission line that capacitively couples to the readout resonator. We down-convert the readout pulse by heterodyne sideband mixing to eliminate dc noise in the Fourier spectrum yielding the dispersive resonator shift. Qubit manipulation is done by heterodyne single sideband mixing. This guarantees that the free evolution of the qubit is not disturbed by microwave leakage through the IQ mixer during pulse off time. In that way other sidebands and the local oscillator (LO) leakage are reduced by at least $40\,\mathrm{dB}$ in power. For $\hat X\hat Y$ control of the qubit, microwave pulses are applied to the same transmission line used for the readout.

The qubit transition frequency is adjusted by a DC current applied to the on-chip flux coil. High frequency noise is filtered at the $4\,\mathrm{K}$ stage with RCR type $\pi$-filters at about $25\,\mathrm{kHz}$ and on the base plate via an RC-element enclosed in copper powder \cite{Lukashenko2008}. Fast flux pulses for $\hat Z$ control of the qubit are sent through a separate microwave line and combined with the offset current by means of a bias tee located at the base plate. The chosen attenuation in the $\hat Z$ pulse line sets the effective electron noise temperature in the center conductor to $\sim 6\,\mathrm{mK}$ above the base temperature.

\section{Parameters of the concentric transmon}

\subsection{System Hamiltonian}

\begin{figure} %++++++++++++++++++++++++++++++++++++++FIG2++++++++++++++++++++++++++++++++++++++++++
\includegraphics{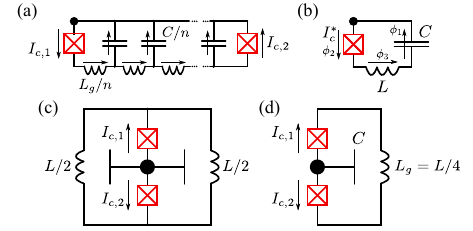}
%\includesvg{eff_schematic05}
\caption{Different schematic circuit diagrams of the concentric transmon qubit. The black dot denotes the central qubit island. (a) Exact lumped element representation of the qubit circuit. (b) Simplified effective circuit diagram. The dc SQUID is treated as a single effective Josephson junction with critical current $I_c ^*$, connected in series with an effective inductance $L$ and the qubit capacitance $C$. It is suggestive to identify the effective inductance with the inductance $L$ of the full outer qubit ring. (c) Geometric representation of the circuit. The qubit capacitance is center symmetrically dispersed. (d) Diagram motivating the choice of effective inductance $L$. Two half rings with inductance $L/2$, respectively, are connected in parallel, forming the geometric inductance $L_g=L/4$.}
\label{fig:eff_schematic}
\end{figure}

The geometric inductance of the large loops in the concentric transmon architecture forms a notable contribution to the total qubit inductance. To account for this, we derive a generalized system Hamiltonian modeling our circuit. The lumped element representation of the qubit circuit is depicted in Fig.~\ref{fig:eff_schematic}(a). The qubit capacitance is distributed center symmetrically into $n$ small parallel capacitances $C/n$ and the total inductance $L_g$ is split into a series of $n$ inductors of inductance $L_g/n$. We approximate this exact circuit by the effective simplified circuit diagram depicted in Fig.~\ref{fig:eff_schematic}(b), consisting of a single capacitor of capacitance $C$ connected in series with an effective inductance $L$ and a single Josephson junction of effective critical current $I_c ^*$, valid in the regime $L\ll \frac{\hbar}{2eI_c}$. The total critical current of the SQUID is treated as an effective parameter $I_c ^*=1.5I_c$ to account for model inaccuracies. The geometric inductance $L_g$ we extract in this analysis is related to the effective inductance $L$ via $L_g=L/4$. This can be motivated since $L_g$ is formed by two half rings in parallel, see Fig.~\ref{fig:eff_schematic}(c), (d). This choice of effective parameters is justified by a good agreement of the presented model with measured data and simulation results. In the following we present a detailed derivation of the system Hamiltonian based on the simplified circuit diagram depicted in Fig.~\ref{fig:eff_schematic}(b).

From current conservation at the two independent active nodes we get
\begin{equation}
\frac{\hbar}{2e}C\ddot{\phi}_1 = I_c ^* \sin\phi_2 = \frac{\hbar}{2e} \frac{1}{L}\phi_3.
\label{eq:current_cons}
\end{equation}
The phases $\phi_i$ in Fig.~\ref{fig:eff_schematic}(b) correspond to the respective voltage drop across each of the components of the circuit according to
\begin{equation}
\frac{\hbar}{2e}\phi_i=\int_{t_0}^t \text{d}t' U_i(t').
\label{eq:U}
\end{equation}
The directed voltage drops in a closed network add up to zero according to Kirchhoff's rule. In a closed loop, the integration constant in Eq.~\eqref{eq:U} is the total magnetic flux penetrating the loop. Since the network in Fig.~\ref{fig:eff_schematic}(d) is an open loop in the sense of the flux quantization law, we can write
\begin{equation}
\phi_1+\phi_2+\phi_3=0.
\label{eq:loop}
\end{equation}
We eliminate $\phi_2$ using Eq.~\eqref{eq:current_cons} and insert in Eq.~\eqref{eq:loop}
\begin{equation}
\phi_3+\arcsin(c\phi_3)=-\phi_1,
\label{eq:phi3_uninverted}
\end{equation}
introducing the notation
\begin{equation}
c=\frac{\hbar}{2e}\frac{1}{I_c ^* L}.
\end{equation}
An approximate solution to Eq.~\eqref{eq:phi3_uninverted} can be written as
\begin{equation}
\phi_3=-\frac{1}{(1+c)^2}\phi_1-\frac{c}{(1+c)^2}\sin\phi_1
\end{equation}
which is exact when $\phi_1\ll 1$ or $c\gg 1$. Plugging into Eq.~\eqref{eq:current_cons} and writing $\phi_1\equiv\phi$, we reduce the set of equations to a single equation of motion:
\begin{equation}
\frac{\hbar}{2e}C\ddot\phi =-\frac{\hbar}{2e}\frac{1}{L} \frac{1}{(1+c)^2}\phi -\frac{\hbar}{2e}\frac{1}{L}\frac{c}{(1+c)^2}\sin\phi
\end{equation}
With the Euler-Lagrange equation
\begin{equation}
\frac{\text{d}}{\text{d}t}\frac{\partial \mathcal{L}}{\partial\left(\frac{\hbar}{2e}\dot\phi\right)} =\frac{\partial \mathcal{L}}{\partial\left(\frac{\hbar}{2e}\phi\right)}
\label{eq:EulerLagrange}
\end{equation}
we obtain the Lagrange function
\begin{eqnarray}
\mathcal{L} & = & \frac{C}{2}\left(\frac{\hbar}{2e}\dot\phi\right)^2 -\frac{1}{2L} \frac{1}{(1+c)^2}\left(\frac{\hbar}{2e}\phi\right)^2\nonumber\\ & & +\left(\frac{\hbar}{2e}\right)^2\frac{1}{L}\frac{c}{(1+c)^2}\cos\phi.
\end{eqnarray}
Employing the definitions of the Josephson energy $E_J=\frac{\hbar}{2e}I_c$, and the inductive energy $E_L=\left(\frac{\hbar}{ 2e}\right)^2\frac{1}{2L_g}$, the parameter $c$ becomes
\begin{equation}
c=\frac{2e}{\hbar}\frac{1}{1.5I_c}\left(\frac{\hbar}{2e}\right)^2\frac{2}{8L_g} =\frac{E_L}{3E_J}.
\end{equation}
Using the conjugated variable for the charge number $n=C\frac{\hbar}{(2e)^2}\dot\phi$, we arrive at the Hamiltonian
\begin{eqnarray}
\hat H & = & 4E_C(\hat n-n_g)^2 -E_J\frac{6E_L^2}{(6E_J+2E_L)^2}\cos\hat\phi\nonumber\\
& & +E_L\frac{9E_J ^2}{(6E_J+2E_L)^2}\hat\phi ^2
\end{eqnarray}
with charging energy $E_C=e^2/2C$.

\subsection{Charging energy $E_C$, Josephson energy $E_J$ and inductive energy $E_L$}

\begin{figure} %++++++++++++++++++++++++++++++++++++++FIG3++++++++++++++++++++++++++++++++++++++++++
\includegraphics{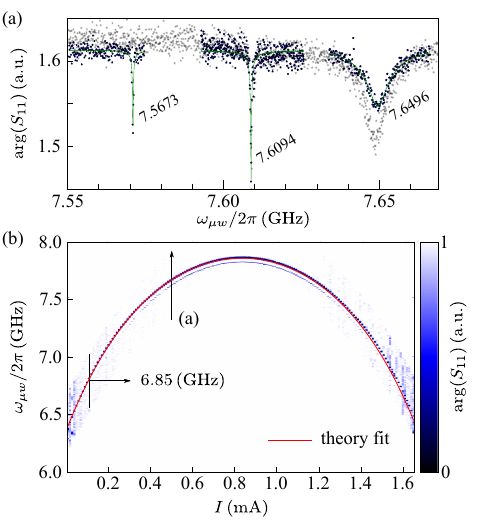}
%\includesvg{spectrum06}
\caption{(a) Fundamental qubit transition and lowest order multi-photon transitions. From the extracted transition frequencies $\omega_{01}/2\pi=7.6496\,\mathrm{GHz}$, $\frac 1 2 \omega_{02}/2\pi=7.6094\,\mathrm{GHz}$ and $\frac 1 3 \omega_{03}/2\pi=7.5673\,\mathrm{GHz}$ we obtain three equations using Eq.~\eqref{eq:E0m}. The transition peaks are measured at varying excitation power (blue data points). Transparent data points correspond to a measurement of the full frequency range. (b) Frequency spectrum of the concentric transmon qubit with respect to the dc current applied to the flux bias line. The thick blue line denotes the measured qubit transition frequency which is fitted (red) to the typical dc SQUID dispersion. The faint blue line is the two-photon transition to the second excited transmon level. Qubit measurements presented in the main text are taken at $\omega_{01}/2\pi=6.85\,\mathrm{GHz}$.}
\label{fig:spectrum}
\end{figure}

Expanding the constructed system Hamiltonian
\begin{equation}
\hat H = 4E_C(\hat n -n_g)^2 -\tilde E _J\cos\hat{\phi}+\tilde E _L\hat\phi ^2,
\label{eq:H}
\end{equation}
with
\begin{equation}
\tilde E _J=E_J\frac{6E_L ^2}{(6E_J+2E_L)^2},\,   \tilde E _L=E_L\frac{9E_J ^2}{(6E_J+2E_L)^2}
\label{eq:tildes}
\end{equation}
up to fourth order in $\hat \phi$ yields
\begin{equation}
\hat H = 4E_C(\hat n -n_g)^2 +\left(\tilde E _J/2+\tilde E _L\right)\hat\phi ^2 -\frac{\tilde E _J}{4!}\hat\phi ^4 +\mathrm{const}.
\label{eq:Hexp}
\end{equation}
We identify $4E_C\equiv\frac {\hbar^2}{2m}$ and $\tilde E _J/2+\tilde E _L\equiv \frac 1 2m\omega^2$ to cast the Hamiltonian in Eq.~\eqref{eq:Hexp} into the standard form of the harmonic oscillator for terms of order $O(\hat\phi ^2$). This yields the phase operator
\begin{equation}
\hat\phi =\left(\frac{E_C}{\tilde E _J/2+\tilde E _L}\right)^{1/4}(\hat a ^\dagger + \hat a)
\end{equation}
in the representation of creation (annihilation) operators $\hat a ^\dagger$ ($\hat a$). Plugging into Eq.~\eqref{eq:Hexp} and taking into account the bosonic commutation relation, $[\hat a,\hat a ^\dagger]=1$, yields
\begin{eqnarray}
\hat H & = & 4\sqrt{E_C(\tilde E _J/2+\tilde E _L)}\hat a ^\dagger \hat a\nonumber\\
& & -\frac{\tilde E _J E_C}{4(\tilde E _J/2+\tilde E _L)}\left((\hat a ^\dagger \hat a)^2+\hat a ^\dagger \hat a\right) +\mathrm{const.}
\label{eq:Harepr}
\end{eqnarray}
and we can find the energies $E_{0m}$ of transmon levels $|m\rangle$ to be
\begin{equation}
E_{0n} =4\sqrt{E_C(\tilde E _J/2+\tilde E _L)}m -\frac{\tilde E _J E_C}{4(\tilde E _J/2+\tilde E _L)}(m^2+m).
\label{eq:E0m}
\end{equation}

\begingroup
\squeezetable
\begin{table} %++++++++++++++++++++++++++++++++++++++TABLE1++++++++++++++++++++++++++++++++++++++++++
\begin{center}
\caption{Extracted parameters of the concentric transmon qubit using data in Fig.~\ref{fig:spectrum}. Errors in $E_C$ and $C$ are assumed, other errors are standard deviations as extracted from the fit. $I_c$ denotes the total critical current of both Josephson junctions. The contribution $L_g$ of the geometric inductance relative to the total qubit inductance $L_{tot}$ at $6.835\,\mathrm{GHz}$ is $L_g/L_{tot}=10\%$.}
\label{tab:params}
\vspace*{1mm}
\begin{ruledtabular}
\begin{tabular}{c|cccccc}
%\begin{tabular}{p{1.35cm}p{1.4cm}p{1.4cm}cp{1.15cm}p{1.15cm}|p{1.3cm}}
& $E_J$ & $I_c$ & $E_C$ & $C$ & $E_L$ & $L_g$ \\
\hline\\
sweet spot & $45\,\mathrm{GHz}$ & $90\,\mathrm{nA}$ & \multirow{2}{*}{$0.24\,\mathrm{GHz}$} & \multirow{2}{*}{$81\,\mathrm{fF}$} & \multirow{2}{*}{$128\,\mathrm{GHz}$} & \multirow{2}{*}{$0.64\,\mathrm{nH}$} \\
$6.85\,\mathrm{GHz}$ & $29\,\mathrm{GHz}$ & $58\,\mathrm{nA}$ \\
\hline\\
$\Delta$ & $\pm 12\,\mathrm{GHz}$ & $\pm 23\,\mathrm{nA}$ & $\pm 0.03\,\mathrm{GHz}$ & $\pm 10\,\mathrm{fF}$ & $\pm 30\,\mathrm{GHz}$ & $\pm 0.16\,\mathrm{nH}$ \\
\end{tabular}
\end{ruledtabular}
\end{center}
\end{table}
\endgroup

Figure~\ref{fig:spectrum}(a) shows the fundamental qubit transition $|0\rangle\leftrightarrow |1\rangle$ and the lowest order multi-photon transitions $1/2\,|0\rangle\leftrightarrow |2\rangle$ and $1/3\,|0\rangle\leftrightarrow |3\rangle$, measured close to the current sweet spot of the spectrum, see arrow in Fig.~\ref{fig:spectrum}(b). Equating Eq.~\eqref{eq:E0m} for $m=1,2,3$ and using measured transition data renders three equations for the transmon parameters $E_C$, $E_J$ and $E_L$. In an iterative approach using this transition data and by fitting measured spectroscopy data, we can extract a suggestive set of parameters which are in good agreement with both data sets. The spectrum fit is carried out using Eq.~\eqref{eq:E0m} evaluated for $m=1$, see Fig.~\ref{fig:spectrum}(b).
$E_J=E_J(\Phi)$ is a function of the applied magnetic flux $\Phi$ with
\begin{widetext}
\begin{equation}
E_J(\Phi)=\frac{\hbar}{2e}I_{c,eff} ^{tot}(\Phi) =\frac{\hbar}{2e}I_c ^{tot} \left|\cos\left(\pi\frac{\Phi}{\Phi_0}+\Delta\Phi\right)\right| \sqrt{1+d^2\tan^2\left(\pi\frac{\Phi}{\Phi_0}+\Delta\Phi\right)}.
\label{eq:EjPhi}
\end{equation}
%\begin{eqnarray}
%& E_J(\Phi)=\frac{\hbar}{2e}I_{c,eff} ^{tot}(\Phi)\nonumber\\
%& =\frac{\hbar}{2e}I_c ^{tot} \left|\cos\left(\pi\frac{\Phi}{\Phi_0}+\Delta\Phi\right)\right| %\sqrt{1+d^2\tan^2\left(\pi\frac{\Phi}{\Phi_0}+\Delta\Phi\right)}.\nonumber\\
%&
%\label{eq:EjPhi}
%\end{eqnarray}
\end{widetext}
The effective critical current $I_{c,eff} ^{tot}(\Phi)$ of both Josephson junctions is dependent on the external magnetic flux $\Phi$, induced by the flux bias line. $I_c ^{tot}$ denotes the total critical current of the concentric transmon at the flux sweet spot, $\Phi=0$ and $\Delta\Phi$ is a phase offset. The square root term in Eq.~\eqref{eq:EjPhi} accounts for a relative asymmetry between the Josephson junctions \cite{Koch2007}, specified by the parameter $d=\frac{I_{c,1}-I_{c,2}}{I_{c,1}+I_{c,2}}$. A separate fit of the qubit spectrum indicates an asymmetry parameter of $d=0.32$, causing a decrease in tunability range of the circuit. The minimum qubit frequency of the investigated sample is $6.3\,\mathrm{GHz}$.

Table~\ref{tab:params} summarizes the parameters extracted from the described fitting algorithm. The qubit capacitance deviates slightly from microwave simulation data of the geometry, predicting $C=58\,\mathrm{fF}$. Possible reasons for this discrepancy are an additional capacitive contribution by the coupling to the readout resonator in the experiment and the additional capacitive effect of the supply lines to the Josephson junctions which are not considered in the simulation. A geometric finite element simulation yields $L_g\sim 0.2\,\mathrm{nH}$, deviating by a factor of three from the fitted value.

We want to point out that the presented model may be considered as a semi-quantitative approach to model the concentric transmon circuit. Within this framework, the assumptions made and the renormalization of parameters is justified by the good agreement of measured data and simulated data with the model.

\section{Finite element simulation of defect loss}

To validate the microscopic analysis of defect loss contributing to qubit decay, we conduct a finite element simulation of the concentric transmon geometry. We model the sample by assuming an oxide layer thickness of $3\,\mathrm{nm}$, formed by $\mathrm{AlO_x}$ ($\varepsilon_r=11.5$), surrounding the central island and the concentric ring electrode of the qubit. For the interface between substrate and aluminum we assume an effective dielectric constant of $\epsilon_r=6$, accounting for the contribution of $\mathrm{SiO_2}$ ($\varepsilon_r=4$) in the interface oxide. The simulation yields an electric field energy fraction of $2.8\cdot 10^{-4}$ residing in the oxide volume. With a phenomenological loss tangent \cite{Paik2010} $\delta_{\mathrm{AlO_x}} =3\cdot 10^{-3}$ of $\mathrm{AlO_x}$, this leads to a total effective loss tangent of $\delta_{TLF} =8.4\cdot 10^{-7}$ attributed to dissipation by surface and interface oxide defects. Due to errors in $\delta_{\mathrm{AlO_x}}$, the oxide layer thickness and $\varepsilon_r$, this value is to be considered exact within a factor of two. $\delta_{TLF}$ can be calculated to limit $T_1$ at $\Gamma_{1,TLF}^{-1}=28\,\mathrm{\mu s}$ which is in good agreement with the microscopically extracted value for $\omega_q/2\pi=6.85\,\mathrm{GHz}$.

In the same spirit we can extract a relative electric field energy ratio in the silicon substrate of $0.92$. With an intrinsic silicon loss tangent $\delta_{Si} < 1\cdot 10^{-7}$, dissipation in the substrate can be neglected. This small value is suggested for instance by very high quality factors measured for TiN resonators on intrinsic silicon \cite{Sandberg2012}.

\bibliography{concentric}